\documentclass[preprintnumbers,amsmath,amssymb,nofootinbib]{revtex4}
\usepackage{amssymb}
\usepackage{latexsym}
\usepackage{color}
\usepackage{graphicx}
\usepackage{hyperref}
\usepackage{color}

\def\nn{\nonumber\\}
\newcommand{\f}[2]{\frac{#1}{#2}}
\def\be{\begin{equation}}
\def\ee{\end{equation}}
\def\bea{\begin{eqnarray}}
\def\eea{\end{eqnarray}}
\begin{document}
\title{Singularity Avoidance in Gravitational Collapse of an Inhomogeneous Fluid in Rastall Gravity}
\author{Akbar Jahan\footnote{jahan@riaam.ac.ir}, Halime Miraghaei \footnote{h.miraghaee@gmail.com}, Shayesteh Ghaffari\footnote{sh.ghaffari@riaam.ac.ir}, Amir Hadi Ziaie\footnote{ah.ziaie@riaam.ac.ir}}
\address{Research Institute for Astronomy and Astrophysics of Maragha (RIAAM), University of Maragheh, P. O. Box 55136-553, Maragheh, Iran}
\begin{abstract}
Various types of inhomogeneous collapse models in general relativity (GR) lead to the formation of spacetime singularities either visible or hidden by a spacetime horizon. Our aim in the present work is to search for nonsingular models in Rastall gravity that arise as the final outcomes of spherically symmetric gravitational collapse of an inhomogeneous matter cloud. We firstly assume linear equations of state (EoS) for radial and tangential pressure profiles, i.e., $p_r=w_r\rho$ and $p_\theta=w_\theta\rho$, then we set the Rastall parameter in such a way that the effective pressure in radial direction vanishes and examine the conditions under which the spacetime singularity can be avoided. We find exact nonsingular collapse solutions for which the collapsing cloud reaches a minimum physical radius at a finite amount of time and then rebounds to an expanding phase where the matter shells start moving away from each other. The solutions we obtain respect the weak energy condition (WEC), which is important for the physical validity of the model.
\end{abstract}

\maketitle
\section{Introduction}
The process of gravitational collapse of a dense body under its own weight is a fundamental issue in gravitation and astrophysics. When a massive star exhausts its nuclear fuel, it can no longer sustain against the pull of gravity and undergoes a continual gravitational collapse. As predicted by Hawking and Penrose singularity theorems~\cite{HAWPENST}, under physically reasonable conditions the end product of such a catastrophic phenomenon is the formation of a spacetime singularity, the event at which, densities as well as spacetime curvatures diverge and the laws of physics as we know them break down~\cite{COUNTERCCC,Joshibook}. In the context of GR, the collapse dynamics is governed by Einstein's field equation, which predicts the formation of an event horizon in the case of black holes, shielding the singularity from external observers~\cite{CCCREF}. However, under some reasonable circumstances interesting collapse scenarios have been reported in the literature where the event horizon may not form leading to the formation of naked singularities which are visible to distant observers see e.g.,~\cite{Joshibook,ColReview} for recent reviews.
\par
The fundamental problems associated with singularity formation in GR, namely path incompleteness and predictability~\cite{problemssin,problemssin1,problemssin2}, have motivated researchers to consider modified theories of gravity and quantum gravity models. In the framework of modified theories of gravity, such as $f(R)$ gravity, scalar-tensor theories, and brane-world scenarios, it is shown that additional degrees of freedom or higher-dimensional effects can alter the dynamics of gravitational collapse and potentially prevent singularity formation~\cite{modfssing} or leading to new types of compact objects~\cite{modfssingcomp}. In the context of scalar-tensor theories, the nonminimal coupling of scalar fields to gravity provide a setting to avoid singularities or modify the formation of event horizons~\cite{stfarsin},\cite{modfssing1}. Brane-world models can significantly alter the collapse process, as gravitational effects in the bulk can influence the dynamics on the brane, potentially hampering singularity formation~\cite{branesing}. From another side, it is widely believed that in very late stages of the collapse process where extreme matter densities and spacetime curvatures are present, the classical GR may no longer hold and a quantum gravity theory, which deals with all forces of nature in a unified way, may take over to resolve the classical spacetime singularity. Hence, quantum gravity approaches, such as loop quantum gravity and string theory may provide a more complete understanding of the final stages of gravitational collapse~\cite{qsingavoid}. These theories suggest that quantum effects may prevent the formation of singularities by introducing a minimum length scale or repulsive forces at high densities~\cite{qsingavoid1}. 
\par
The issue of singularity avoidance is a major field of research in the study of a typical collapse process. As mentioned above, nonsingular collapse models can be probed in the classical framework of alternative gravity theories as well as quantum gravitational models. Many investigations in this direction have been carried out and the results reveal that these theories often give rise to a nonsingular state where the collapse halts before reaching infinite curvature and density, consequently leading to stable remnants or other exotic outcomes~\cite{exocol}. Additionally, bouncing scenarios in quantum collapse models have attracted much attention, proposing that collapsing matter cloud may rebound after reaching a critical density, avoiding thus the singularity formation\cite{quantuminhom}~\cite{qcolbounce}. Regarding the above considerations, within the framework of modified gravity theories, it is of particular interest to investigate the possibility of singularity avoidance in the collapse process or cosmological scenarios. Work along this line has been extensively performed in the literature, see e.g.,~\cite{modfssing,modfssingcomp,modfssing1,stfarsin,branesing},~\cite{nonsingcos}.
\par
A rather simple and physically meaningful way to modify GR is to consider a non-minimal coupling between matter and geometry. Such an assumption leads to violation of the ordinary energy-momentum conservation law, i.e, $\nabla_\mu {\rm T}^\mu_{\,\,\,\, \nu}=0$~\cite{od1,od2,cmc,cmc1,cmc2,rastall}. This idea which was firstly proposed by Peter Rastall~\cite{rastall} states that, the ordinary conservation law of the energy-momentum tensor (EMT) should be modified to $\nabla^\mu {\rm T}_{\mu\nu}=\lambda\partial_\nu{\mathcal R}$ where ${\mathcal R}$ is the Ricci curvature scalar and $\lambda$ is a constant which measures the strength of non-minimal coupling between the geometry and matter sources. The main physical motivations for Rastall theory can be found in particle production process in cosmology~\cite{pppcos} and that, the conservation laws have been only tested on the Minkowski background spacetime or in the presence of quasistatic gravitational fields~~\cite{rastall},\cite{qusgrf} and hence, they may no longer hold in curved spacetime. Since its advent, the Rastall theory has received significant attentions as it is in good agreement with various observational data and theoretical expectations~\cite{genras,dmy,FESRASPLB,Effemtras,validity1,validity2}. The collapse process of a homogeneous perfect fluid in Rastall gravity has been addressed in~\cite{colrast} where the authors found singular solutions to Rastall field equation
for which the spacetime evolves to a final state that is either a black hole or a naked singularity, depending on the values of Rastall and EoS parameters. Also gravitational collapse of a null fluid in Rastall theory has been investigated in~\cite{ziataval}. More interestingly, nonsingular outcomes as corrections/alternatives to singular homogeneous dust collapse scenario, known as Oppenheimer-Snyder-Datt (OSD) model~\cite{OSDM}, were obtained and discussed in~\cite{colrast1}. In this work the authors assumed a varying coupling parameter, and showed that the singularity that occurs in the OSD model is replaced by a nonsingular bounce event. Nevertheless, over the past few years, there have been some criticisms against Rastall gravity in the literature which argue that Rastall theory is fundamentally equivalent to GR~\cite{eqrasgr}. This issue has raised counterarguments in favor of this theory which can be summarized as follows: $i)$ While the field equation of Rastall gravity may be mathematically recast into GR-like form, the physical meaning of the EMT in Rastall theory differs and indeed, the non-conservation of EMT in Rastall theory implies a non-minimal coupling between matter and geometry, which is not present in GR~\cite{dmy}. This could lead to different interpretations of cosmological or astrophysical phenomena~\cite{FESRASPLB,Effemtras,validity1,validity2},\cite{validras}. $ii)$ Rastall gravity explicitly violates the standard conservation law which is the cornerstone of GR. Such violation has implications for thermodynamics, black hole physics, gravitational lensing effects and quantum field theory in curved spacetime~\cite{rastall,pppcos,qusgrf},\cite{glensras}, see also~\cite{dmy} and references there in. $iii)$ Rastall theory can deviate in cosmological scenarios with perfect fluid as the matter content, for example, the equations of state of dark matter or dark energy may differ in Rastall framework~\cite{fluidras}. Also, stellar structure solutions such as neutron stars could exhibit deviations due to the modified conservation law~\cite{olive2015}.
\par
In the study of gravitational collapse, inhomogeneous models are of particular importance where the collapsing matter distribution is not uniform. The inhomogeneous models provide a more realistic setting for investigating gravitational collapse, as in these models in addition to the time dependence, the density and pressure profiles of the matter can change as a function of the coordinate radius $r$~\cite{singhjoshi1994}. The collapse scenario in these models often lead to the formation of a shell-focusing singularity which is a genuine spacetime singularity where all matter shells crush to a zero physical radius with singular epoch being a
function of coordinate radius, as a result of the inhomogeneity in the matter distribution~\cite{Joshibook}. Over the past decades, much research has been conducted in this field and the results show that under certain conditions, these singularities can be naked, violating thus the cosmic censorship conjecture (CCC)~\cite{CCCREF} and posing serious challenges to the issue of predictability and determinism of GR~\cite{COUNTERCCC}~\cite{JoshiDw1993,JoshiDw19930},\cite{JoshiDw19931,sczsing}. One then may be motivated to seek for inhomogeneous collapse models for which, quantum gravity corrections alter the collapse dynamics and consequently the classical central singularity is replaced by a nonsingular bounce event~\cite{quantuminhom}~\cite{quantuminhom1}. Also, inhomogeneous collapse in the framework of Einstein-Cartan theory has been studied in~\cite{luzmenazia} where it is shown that the effects of intrinsic angular momentum (spin) of matter could provide a setting to avoid singularity formation. In the present work, we are motivated to consider a simple model for gravitational collapse of an inhomogeneous fluid in Rastall gravity and study under what conditions the spacetime singularity can be removed. The organization of this paper is then as follows: In Sec.~(\ref{sec2}) we give a brief review on Rastall theory and then proceed to find the field equations that govern the collapse dynamics. In Sec.~(\ref{sec3}) we try to find exact solutions representing nonsingular collapse models and our conclusions are given in Sec.~(\ref{sec4}).
\section{Field equations and the collapse setting}\label{sec2}
In the framework of Rastall gravity~\cite{rastall} it is proposed that the conventional conservation law for matter EMT is actually questionable in curved spacetime. The modified conservation law introduced by Rastall is formulated as $\nabla_{a}{\rm T}^{a}_{\,\,\,b}=\lambda\nabla_{b}{\mathcal R},$
where $\lambda$ is the Rastall parameter. Such a generalization on EMT conservation leads to the modification of Einstein tensor as~\cite{rastall,FESRASPLB}
\be\label{Einsmod}
G_{ab} \rightarrow G_{ab}+\kappa_{r}\lambda g_{ab}{\mathcal R},
\ee
where $\kappa_{r}$ is a constant. The Rastall field equation then reads
\be\label{RastallFES}
{\rm G}_{ab} +\gamma g_{ab}{\mathcal R}=\kappa {\rm T}_{ab},
\ee
where $\kappa=8\pi G/c^4$ is the Einstein's gravitational constant and $\gamma=\kappa_r\lambda$ is the Rastall parameter. Moreover, as it was argued by Rastall~\cite{rastall} in the weak field limit one can verify that $\kappa_r=(1-4\gamma)\kappa/(1-6\gamma)$ which shows that the GR field equation, i.e., $\kappa_r\rightarrow\kappa$ will be recovered in the limit $\lambda\rightarrow0$~\cite{weakfield}. Taking the trace of Eq.~(\ref{RastallFES}) and substituting the result for the Ricci scalar we arrive at the equivalent form of the field equation in terms of the effective source as
\be\label{FESEquiv}
{\rm G}_{ab}=\kappa_{r}{\rm T}^{\rm eff}_{ab},~~~~~~~~~{\rm T}^{\rm eff}_{ab}={\rm T}_{ab}-\f{\gamma g_{ab}{\rm T}}{4\gamma-1}.
\ee 
From the above equation we can deduce that there appears a nonminimal coupling between geometry (spacetime metric) and matter fields (trace of EMT). The Rastall parameter acts as a measure of such coupling which is usually understood as mutual interaction between geometry and matter~\cite{od1,od2,cmc,cmc1,cmc2,rastall}~\cite{weakfield}. The general spherically symmetric line element describing the collapsing matter cloud in comoving coordinates is given by
\be\label{met}
ds^2=-e^{2\nu(t,r)}dt^2+e^{2\psi(t,r)}dr^2+R^2(t,r)d\Omega^2,
\ee
where $d\Omega^2$ is the standard line element on the unit two-sphere and $\nu$, $\psi$ and $R$ are functions of $t$ and $r$. We note that the function $R$ with condition $R(t,r)\geq0$ is the area coordinate in such a way that the quantity $4\pi R^2(t,r)$ is equivalent to the proper area of the mass shells and the area of such a shell with $r=cte.$ vanishes when $R(t,r)\rightarrow0$. Therefore, the curve $R(t,r)=0$ describes the spacetime singularity, an event at which all the mass shells with $0\leq r<\infty$ collapse to a vanishing volume. This type of singularity is usually known as a shell-focusing singularity~\cite{Joshibook,COUNTERCCC}. For spherically symmetric spacetime considered here we assume that the collapsing matter is of the so-called type (I) matter fields which include all the known physical forms of matter, such as dust, perfect fluids, and massless scalar fields~\cite{HAWPENST}. This class of matter is specified by requiring that the corresponding EMT admits one timelike and three spacelike eigenvectors~\cite{HAWPENST},\cite{Stef2003}, except the special cases that are classified as type (II) sources (null fluids) for which the corresponding EMT admits a double null eigenvector~\cite{HAWPENST}. The class of spacetimes that are natural candidates for describing type (II) matter fields are the generalized Vaidya spacetimes~\cite{Wang1999}. In the background of such spacetimes, the study of collapse scenario and its final outcome has been extensively addressed in the literature, see e.g.,~\cite{ziataval},\cite{nakedvaidya}. The EMT components of a type (I) matter in comoving frame can be written in diagonal form as
\bea\label{emtcomps} 
{\rm T}^{t}_{\,t}=-\rho(t,r),~~~~~{\rm T}^{r}_{\,r}=p_r(t,r),~~~~~{\rm T}^{\theta}_{\,\theta}=p_\theta(t,r)={\rm T}^{\phi}_{\,\phi}=p_\phi(t,r),~~~~~{\rm T}^{t}_{\,r}={\rm T}^{r}_{\,t}=0,
\eea
where $\rho$, $p_r$ and $p_\theta$ are the density, radial and tangential pressures, respectively. For the sake of physical reasonability, the collapsing matter is usually taken to satisfy the WEC, i.e., the energy density as measured by an observer is non-negative and for any timelike vector field $u^a$, the inequality ${\rm T}_{ab}u^au^b\geq0$ holds. For a type (I) matter the WEC is fulfilled provided that
\be\label{WEC}
\rho\geq0,~~~~~~\rho+p_r\geq0,~~~~~~\rho+p_\theta\geq0.
\ee
Using the second part of Eq.~(\ref{FESEquiv}), the components of effective EMT are obtained as
\bea
{\rm T}^{t\,{\rm eff}}_{\,\,t}\equiv-\rho^{\rm eff}&=&-\f{(3\gamma-1)\rho+\gamma(p_r+2p_t)}{4\gamma-1},\label{EFFEMT1}\\
{\rm T}^{r\,{\rm eff}}_{\,\,r}\equiv p_r^{\rm eff}&=&\f{(3\gamma-1)p_r+\gamma(\rho-2p_t)}{4\gamma-1},\label{EFFEMT2}\\
{\rm T}^{\theta\,{\rm eff}}_{\,\,\theta}={\rm T}^{\phi\,{\rm eff}}_{\,\,\phi}&\equiv& p_\theta^{\rm eff}=\f{(2\gamma-1)p_t+\gamma(\rho-p_r)}{4\gamma-1},\label{EFFEMT3}\\
{\rm T}^{r\,{\rm eff}}_{\,\,t}&=&{\rm T}^{t\,{\rm eff}}_{\,\,r}=0.\label{EFFEMT4}
\eea
We also require that the components of effective EMT satisfy the WEC hence we have
\be\label{effemtwec}
\rho^{\rm eff}\geq0,~~~~~~~~\rho^{\rm eff}+p^{\rm eff}_r\geq0,~~~~~~~~\rho^{\rm eff}+p^{\rm eff}_\theta\geq0.
\ee
For the spacetime metric (\ref{met}) the field equation (\ref{FESEquiv}) along with Bianchi identity $\nabla_a{\rm T}^{a\,{\rm eff}}_b=0$ lead to the following set of differential equations (We have set the units so that $\kappa=1$)
\bea
&&\f{F^\prime}{R^2R^\prime}=\kappa_r\rho^{\rm eff},~~~~~~~~~\f{\dot{F}}{R^2\dot{R}}=-\kappa_rp_r^{\rm eff},\label{eqsfmass}\\
&&\dot{R}^\prime-\dot{\psi}R^\prime-\nu^\prime\dot{R}=0,\label{fesbia3}\\
&&p_r^{\prime\,\rm eff}+\left[{\nu}^\prime+\f{2{R}^\prime}{R}\right]p_r^{\rm eff}+{\nu}^\prime \rho^{\rm eff}=2p^{\rm eff}_\theta\f{{R}^\prime}{R},\label{fesbia5}\\
&&F=F(t,r)=R\left(1-{\rm e}^{-2\psi}R^{\prime2}+{\rm e}^{-2\nu}\dot{R}^{2}\right),\label{massdef}
\eea
where $\prime\equiv\partial/\partial r$, $\dot{}\equiv\partial/\partial t$. The function $F(t,r)$ is defined as the mass function of the system and is physically interpreted as the amount of matter enclosed in the comoving shell labeled by $r$ at time $t$~\cite{MSE}. In order to find analytical solutions for this system we firstly determine the EoS in radial and tangential directions. A linear EoS is well motivated by several theoretical and practical considerations. However, one may choose more complicated forms such as polytropic EoS $p=K\rho^{1+\f{1}{n}}$ with $K$ being the polytropic constant and $n$ being the polytropic index~\cite{polyeos,polyeos1}, quadratic EoS $p=\alpha\rho+\beta\rho^2$~\cite{quadeos}, Chaplygin $p=-A/\rho$ and generalized Chaplygin gas $p=-A/\rho^\alpha$~\cite{polchapeos}, Van der Waals $p=\gamma\rho/(1-\beta\rho)-\alpha\rho^2$~\cite{vandereos} and Logarithmic EoS $p=A\ln(\rho/\rho_0)$~\cite{logeos}. The first reason for adopting a linear EoS is that, it provides a mathematically tractable framework that allows for exact analytical solutions to the complicated set of the field equations, while models with nonlinear EoS would often require numerical considerations. The second reason is that, a linear EoS captures essential phenomenological behaviors therefore enables direct comparisons with known matter fields and energy conditions. For example the EoS parameters for dust and radiation fluids are specified by $w=0$ and $w=1/3$, respectively. In $\Lambda{\rm CDM}$ model, a cosmological constant represents the dark energy with EoS parameter $w\approx-1$~\cite{lucadarkeos},\cite{vangoozi}. In canonical scalar field models the quintessence field is described by the EoS parameter $-2/3\leq w\leq-1/3$~\cite{quinteeos} and phantom fields are characterized by $w<-1$~\cite{lucadarkeos},\cite{phantomeos}. We therefore assume the fluid components obey linear EoS, i.e., $p_r=w_r\rho$ and $p_\theta=w_\theta\rho$, where the EoS parameters $(w_r,w_\theta)\in\mathbb{R}$ are two dimensionless quantities which characterize the state of the fluid in terms of its energy density and pressure profiles. The non-vanishing components of the effective EMT then read
\bea
\rho^{\rm eff}&=&\f{(w_r+2w_\theta+3)\gamma-1}{4\gamma-1}\rho,\label{EFFEMTw1}\\
p_r^{\rm eff}&=&\f{(3w_r-2w_\theta+1)\gamma-w_r}{4\gamma-1}\rho,\label{EFFEMTw2}\\
p_\theta^{\rm eff}&=&\f{(2w_\theta-w_r+1)\gamma-w_\theta}{4\gamma-1}\rho.\label{EFFEMTw3}
\eea
Next, we assume effective pressure in the collapsing configuration is zero, that is, $p_r^{\rm eff}=0$. This can be achieved once we set the Rastall parameter as
\be\label{Rasdust}
\gamma=\f{w_r}{3w_r-2w_\theta+1}.
\ee
We note that the Rastall parameter\footnote{Rastall theory has been extensively investigated to determine its consistency with observational data, particularly in cosmology and astrophysics~\cite{genras,dmy,FESRASPLB,Effemtras,validity1,validity2}. Observational constraints on the Rastall parameter often arise from cosmological probes such as supernovae Type Ia, cosmic microwave background anisotropies, galaxy-scale strong gravitational lensing and baryon acoustic oscillations~\cite{validras},~\cite{observgama}. Studies suggest that Rastall gravity can mimic dark energy effects, providing an alternative explanation for the accelerated expansion of the universe without invoking a cosmological constant~\cite{genras},\cite{darkrastall}.} governs how strongly the changes in energy-momentum distribution influences the spacetime curvature and vice versa. Therefore, in Rastall framework, if we consider the non-minimal coupling between geometry and matter fields as a \lq{}\lq{}{\it degree of freedom}\rq{}\rq{} (which is absent in GR), then, nontrivial behaviors in fluid components is expected, see e.g.~\cite{thermoras} for the effects of such a coupling on the thermodynamic pressure of black holes. Consequently, one may set the Rastall coupling parameter in such a way that the collapsing matter effectively behaves like a dust fluid. This scenario remarkably simplifies the calculations and provides a suitable setting to search for exact inhomogeneous collapse solutions with regular outcomes. Now, from the second part of Eq.~(\ref{eqsfmass}) we readily find $F(t,r)=F(r)$. From Eq.~(\ref{fesbia5}) we can solve for the metric function $\nu$ with the solution given as
\be\label{nusol}
\nu(t,r)=\ln R(t,r)^{\f{2(w_\theta-w_r)}{w_r+1}}+F_1(t),
\ee
where $F_1(t)$ is an arbitrary function of comoving time. Substituting the above result into Eq.~(\ref{fesbia3}) and solving for the metric function $\psi$ we get
\be\label{psisol}
\psi(t,r)=\ln \left[R^\prime(t,r)R(t,r)^{\f{2(w_r-w_\theta)}{w_r+1}}\right]+F_2(r),
\ee
where again, $F_2(r)$ is an arbitrary function of radial coordinate. The remaining equations i.e., first part of Eq.~(\ref{eqsfmass}) and Eq.~(\ref{massdef}) provide us with the matter distribution and collapse dynamics, respectively. Let us now rescale the area radius function $R(t,r)$ so that at initial time $t=0$ at which the collapse begins, we have $R(0,r)=r$. Therefore, the first part of Eq.~(\ref{eqsfmass}) leaves us with the following integral for the mass function
\be\label{Frint}
F(r)=\kappa_{r}\int_{0}^{r}z^2\rho^{\rm eff}(0,z)dz.
\ee
Once the initial density of the collapsing body is specified the mass function is determined. Finally, using Eq.~(\ref{massdef}) along with considering solutions (\ref{nusol}) and (\ref{psisol}) we arrive at the master equation that governs the dynamics of the area radius, given as
\be\label{Reqdyn}
F(r)-R(t,r)-R(t,r)^\beta\dot{R}(t,r)^2f_1(t)+R(t,r)^\delta f_2(r)=0,
\ee
where
\bea\label{betadelta}
\beta=\f{5w_r-4w_\theta+1}{w_r+1},~~~~~~\delta=\f{4w_\theta-3w_r+1}{w_r+1},~~~~~F_1(t)=-\f{1}{2}\ln f_1(t),~~~~~F_2(r)=-\f{1}{2}\ln f_2(r).
\eea
Our aim in the next section is to search for a class of exact solutions to Eq.~(\ref{Reqdyn}) and examine whether a shell-focusing singularity can be prevented.
\section{Collapse Dynamics and Singularity Avoidance}\label{sec3}
Let us consider the following relation between the EoS parameters
\be\label{eoss2}
w_r=\f{4}{3}w_\theta+\f{1}{3},
\ee
whence Eq.~(\ref{Reqdyn}) reads
\be\label{eqclass2}
F(r)+f_2(r)-R(t,r)\left[1+f_1(t)R(t,r)\dot{R}^2(t,r)\right]=0.
\ee
The solution to the above differential equation is given by
\bea\label{solclass2}
R_\pm(t,r)=\f{1}{2}{\mathcal G}_\pm(t,r)^{\f{1}{3}}+\f{2[F(r)+f_2(r)]^2}{{\mathcal G}_\pm(t,r)^{\f{1}{3}}}-f_2(r)-F(r),
\eea
where 
\bea\label{mathgtr}
{\mathcal G}_\pm(t,r)&=&3\left[\left(h(r)\pm\int_{1}^{t}\f{dx}{\sqrt{f_1(x)}}\right)^2\left(9\left(h(r)\pm\int_{1}^{t}\f{dx}{\sqrt{f_1(x)}}\right)^2-16[F(r)+f_2(r)]^3\right)\right]^{\f{1}{2}}\nn&+&8[F(r)+f_2(r)]^3-9\left[h(r)\pm\int_{1}^{t}\f{dx}{\sqrt{f_1(x)}}\right]^2,
\eea
and $h(r)$ being an arbitrary function of radial coordinate, which can be determined using the rescaling $R(0,r)=r$. By doing so, for each sign of the above solution we get two solutions for this function given by
\bea\label{hfuncs}
h_1^\pm(r)&=&\pm\f{2}{3}\left[2f_2(r)+2F(r)+r\right]\sqrt{f_2(r)+F(r)-r}-\int_{0}^{1}\f{dx}{\sqrt{f_1(x)}},\nn
h_2^\pm(r)&=&\pm\f{2}{3}\left[2f_2(r)+2F(r)+r\right]\sqrt{f_2(r)+F(r)-r}+\int_{0}^{1}\f{dx}{\sqrt{f_1(x)}}.
\eea
Substituting for $h_1^\pm(r)$ and $h_2^\pm(r)$ into Eq.~(\ref{mathgtr}) and after simplification, we arrive at four solutions for the area radius that only the two of which are independent. These two solutions are given in general in terms of Eq.~(\ref{solclass2}) with the following substitution
\bea\label{sol1R}
{\mathcal G}_\pm(t,r)\rightarrow {\mathcal Z}_\pm(t,r)&=&3\left[\left(\int_{0}^{t}\f{dx}{\sqrt{f_1(x)}}\pm U(r)\right)^2\left(9\left(\int_{0}^{t}\f{dx}{\sqrt{f_1(x)}}\pm U(r)\right)^2-16[F(r)+f_2(r)]^3\right)\right]^{\f{1}{2}}\nn&+&8[F(r)+f_2(r)]^3-9\left[\int_{0}^{t}\f{dx}{\sqrt{f_1(x)}}\pm U(r)\right]^2,
\eea
where
\be\label{sfuncr}
U(r)=\f{2}{3}\left[2F(r)+2f_2(r)+r\right]\sqrt{F(r)+f_2(r)-r}.
\ee
\par
\begin{figure}
	\begin{center}
		\includegraphics[width=7.7cm]{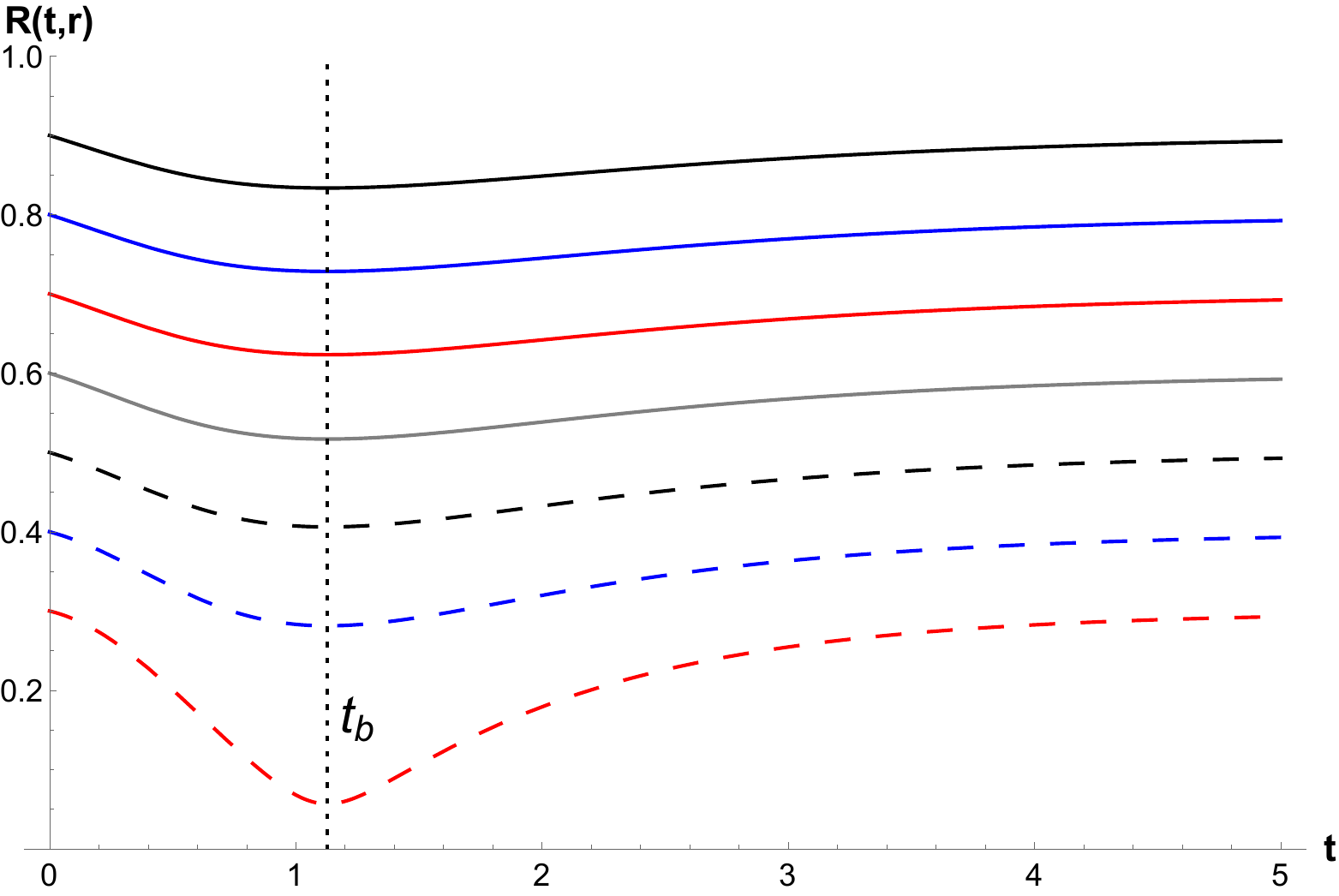}
		\includegraphics[width=7.7cm]{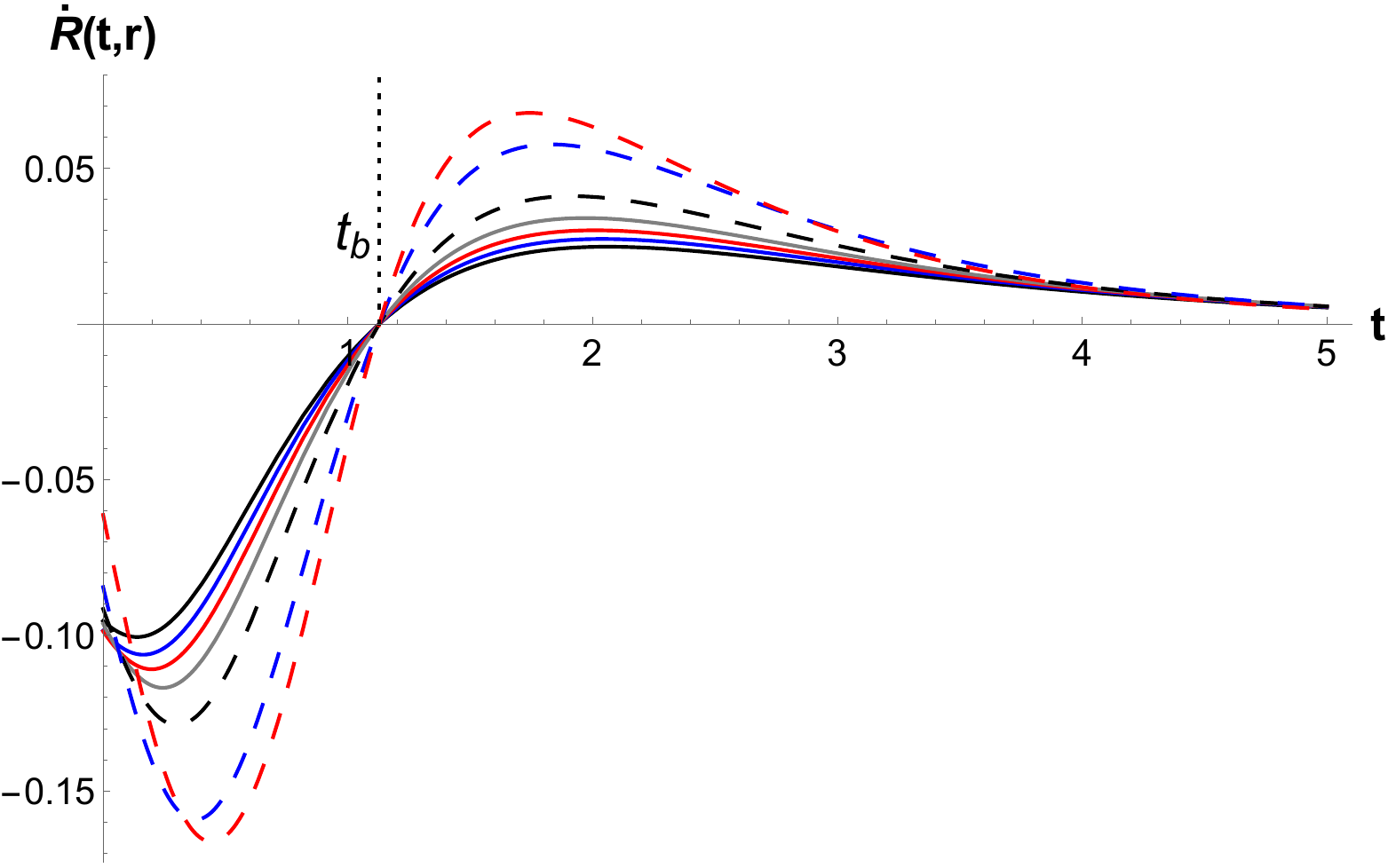}
		\caption{Evolution of area radius (left panel) and the speed of collapse (right panel) for each matter shell. In the left panel, the radial coordinates which label each shell has been chosen as $r=0.9,0.8,0.7,0.6,0.5,0.4,0.3$ for the curves from up down. The model parameters has been set as $n=2$, $w_\theta=0.45$, $\rho_0=1$, $\alpha=10$, $\beta=-1$, $\xi=1$, $\zeta=0.001$, $\delta=0.0001$, $r_{\rm b}=1$ and $\omega=1.83$. The same model parameters have been taken for the right panel. For the chosen model parameters the bounce occurs at $t_{\rm b}=1.1286$ which is shown by the vertical dotted line.}\label{fig1}
	\end{center}
\end{figure}
\begin{figure}
	\begin{center}
		\includegraphics[width=7.7cm]{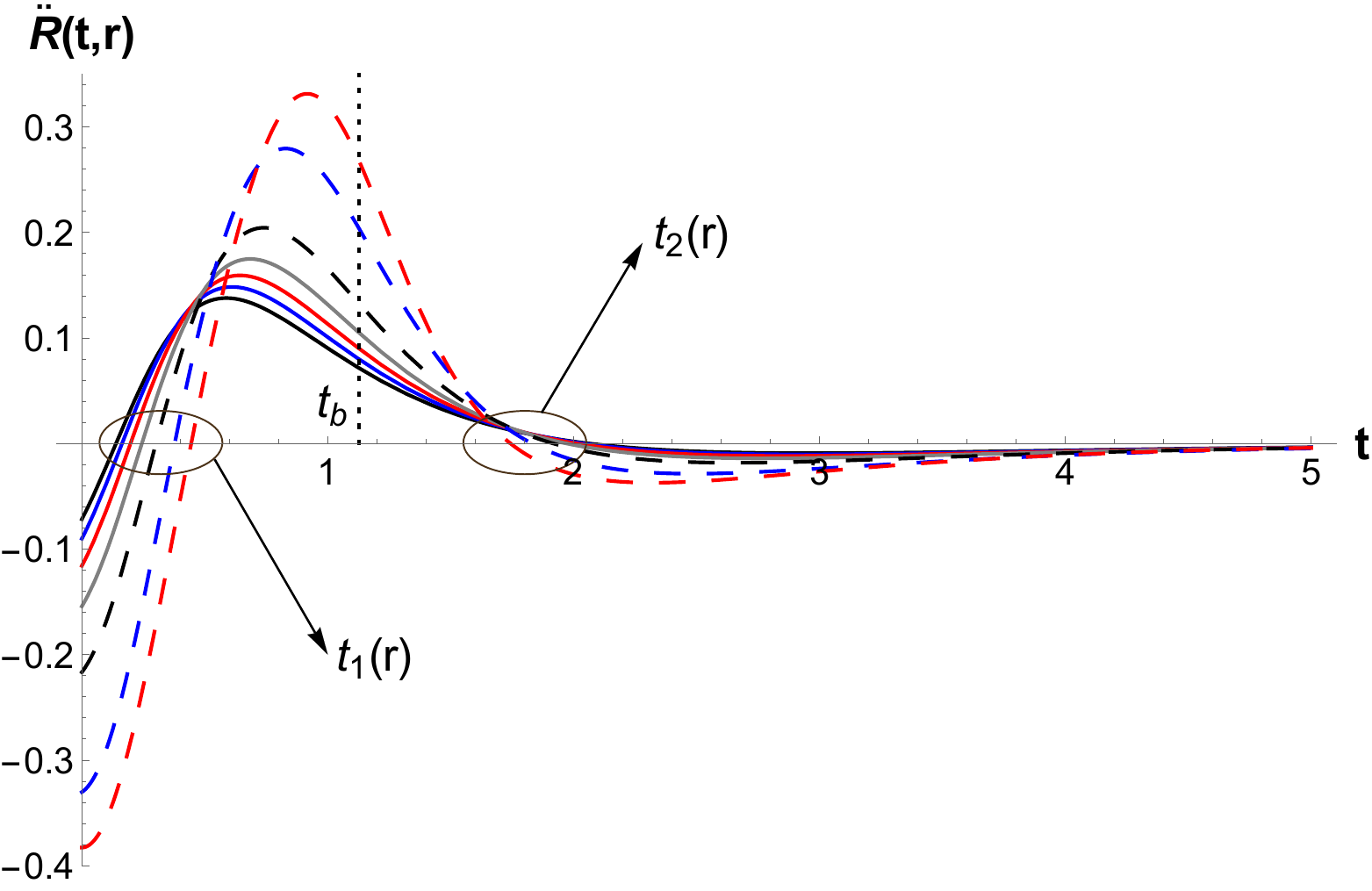}
		\includegraphics[width=7.7cm]{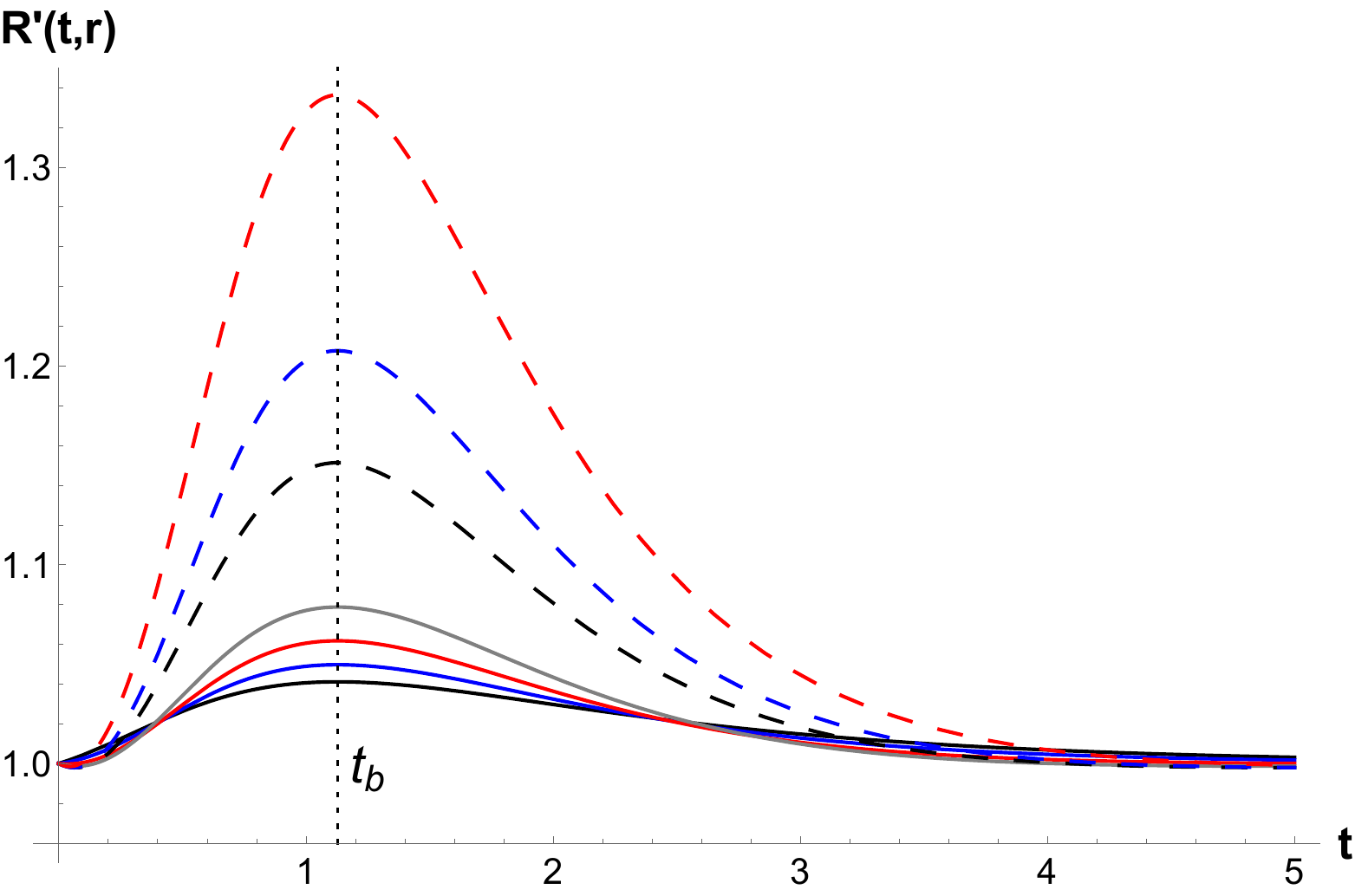}
		\caption{Evolution of collapse acceleration (left panel) and spatial derivative of the physical radius (right panel) for the same values of model parameters as of Fig.~(\ref{fig1}).}\label{fig2}
	\end{center}
\end{figure}

Next, we proceed to determine the mass function $F(r)$ along with the two free functions $f_1(x)$ and $f_2(r)$ in order to extract the final relation for the physical radius of the collapsing body. Let us begin with finding the mass function which can be obtained by performing the integration given in Eq.~(\ref{Frint}). We now define the initial energy density of the cloud as 
\be\label{initialen}
\epsilon(r)=\rho_0\left[1-\left(\f{r}{r_{\rm b}}\right)^n\right],
\ee
where $n\in\mathbb{R}^+$ and $r_{\rm b}$ is the coordinate radius of the outermost shell of matter for which the physical area radius is given by, $R(0,r_{\rm b})=r_{\rm b}$. The mass function then reads
\be\label{massfunc}
F(r)=\f{4\rho_0r^3(w_\theta+1)(5w_\theta-1)}{81(n+3)w_\theta}\left[n+3\left(1-\f{r^n}{r_{\rm b}^n}\right)\right],~~~~~~w_\theta\neq0.
\ee
We note that the mass function vanishes at the center of the cloud $F(0)=0$ and is a positive increasing function of radial coordinate for $w_\theta>1/5$. A suitable choose of the free functions leads us to physically reasonable regular collapse solutions, i.e., those for which the spacetime singularity that may occur at final stages of the collapse scenario, can be replaced by a non-singular bounce. In order to obtain such solutions we propose the following forms for the free functions
\bea\label{freefs}
f_2(r)=\delta+\f{\xi}{r^\beta+\zeta},~~~~~~~f_1(x)=\f{4\alpha\cosh(\omega  x)}{\left[2-\omega  x\tanh(\omega  x)\right]^2}~~\Rightarrow~~\int_{0}^{t}\f{dx}{\sqrt{f_1(x)}}=\frac{t}{\sqrt{\alpha  \cosh (\omega  t)}},
\eea
where the parameters $\{\alpha,\beta,\delta,\zeta,\xi,\omega\}\in\mathbb{R}$. Substituting for the mass and free functions into Eq.~(\ref{sol1R}) and choosing the plus sign\footnote{The minus sign, to our knowledge, represents expanding solutions which can be utilized in the inhomogeneous cosmological models~\cite{inhomcos}. Therefore, a suitable choose of the free functions may help to prevent the initial singularity of the Universe.}, we finally arrive at the desired solution for evolution of the collapsing object. The left panel in Fig.~(\ref{fig1}) shows the behavior of area radius as a function of comoving time for different values of the shell coordinate radii. We observe that the physical area radius decreases monotonically and reaches a minimum value where the bounce event occurs at $t=t_{\rm b}$. It then increases in the post-bounce regime ($t>t_{\rm b}$) until reaching a finite nonzero value. For the present solutions the bounce time is same for all values of $r$ coordinate which means that all the mass shells reach at their corresponding minimum area radius simultaneously. This behavior can be verified through the right panel where we have sketched the speed of collapse against comoving time. It is therefore seen that for all the mass shells we have $\dot{R}(t_{\rm b},r)=0$. The occurrence of a minimum and a maximum in the $\dot{R}(t,r)$ curve at contracting $t<t_{\rm b}$ and expanding $t>t_{\rm b}$ regimes may provide us with more information on the dynamical evolution of the collapsing body. 
To have a better understanding of this issue we consider the acceleration of the collapse scenario which is given by $\ddot{R}(t,r)$ curve. The left panel in Fig.~(\ref{fig2}) shows the behavior of collapse acceleration where we observe that this quantity vanishes at two inflection points where $\ddot{R}(t_1(r),r)=0$ and $\ddot{R}(t_2(r),r)=0$. For $t<t_1(r)$, the matter shells are in an accelerated contracting regime where $\ddot{R}(t,r)<0$ and $\dot{R}(t,r)<0$. When the first inflection point is passed, the shells enter a decelerated contracting regime where $\ddot{R}(t,r)>0$ and $\dot{R}(t,r)<0$. This regime holds until the bounce time is reached after which, for $t>t_{\rm b}$ an accelerated expanding regime commences where $\ddot{R}(t,r)>0$ and $\dot{R}(t,r)>0$. The expanding evolution of the shells continue in this regime until the second inflection point is reached. Then, for $t>t_2(r)$ the matter shells undergo a decelerated expanding phase where $\ddot{R}(t,r)<0$ and $\dot{R}(t,r)>0$. As the time passes, the speed of collapse as well as its acceleration tend to zero values, that is, the collapsing object has achieved an equilibrium state with a constant non-vanishing area radius. It should be noted that the inflection times are different for each shell radius. From the above considerations we can figure out that the evolution of the collapse process is such that the area radius never vanishes hence the collapse scenario does not lead to a shell-focusing singularity at which the entire collapsing cloud crushes to a zero physical radius at the singular event. The formation of such singularity which is a true curvature singularity of the spacetime has been widely reported in different inhomogeneous collapse models, see e.g.,~\cite{Joshibook},\cite{ColReview} and references therein. However, during the dynamical evolution of an inhomogeneous collapse setting, it is still possible for the existence of another type of singular regions, the so-called shell-crossing singularities~\cite{Joshibook},\cite{shellcr1}. At these events, the nearby matter shells intersect and cross each other producing temporary divergences in energy density and curvature scalars~\cite{shellcr}. In contrast to strong curvature shell-focusing singularities such as the Big-Bang and the curvature singularity at the center of the Schwarzschild spacetime~\cite{problemssin1},\cite{ellissing}, the shell-crossing singularities are usually recognized as gravitationally weak singularities which can be resolved by a suitable extension of the spacetime through a coordinate system transformation~\cite{shellcr2,shellcr21}. In order to check whether shell-crossings are present within the evolving matter cloud we may consider the behavior of spatial derivative of the physical radius, i.e., $R^\prime(t,r)$. The right panel of Fig.~(\ref{fig2}) shows the behavior of this quantity where we observe that for $t<t_{\rm b}$, the matter shells start moving away from each other to a maximum distance where $\dot{R}^\prime(t_{\rm b},r)=0$. However, this does not mean that the shells are expanding as the area radius is a decreasing function of comoving time. Indeed, while the entire cloud undergoes a collapse process, the inner shells falls into a nonsingular center with larger acceleration than the outer ones, hence, the relative distance between the shells increases until the bounce event arrives. For the present model, the matter shells reach the maximum distance at the bounce time, after which, they enter an expanding phase where again, the inner shells start to expand in a faster rate than the outer ones and compensate the difference in relative distance that occurred in the contracting regime. This behavior can be understood from the  accelerated contracting and decelerated expanding regimes. We therefor conclude that the matter shells never meet each other ($R^\prime\neq0$) and hence no shell-crossings take place.
\par
\begin{figure}
	\begin{center}
		\includegraphics[width=7.7cm]{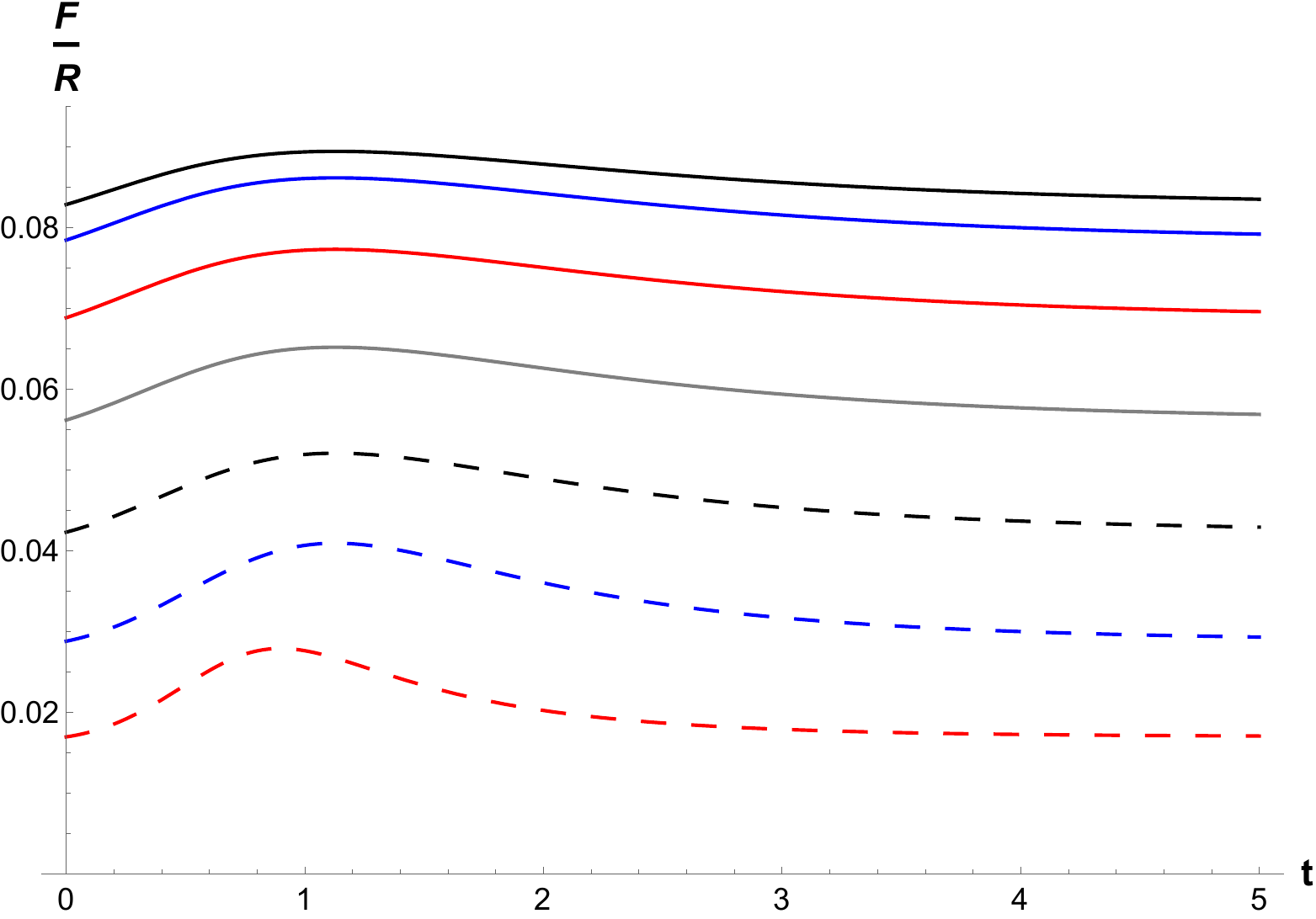}
		\includegraphics[width=7.1cm]{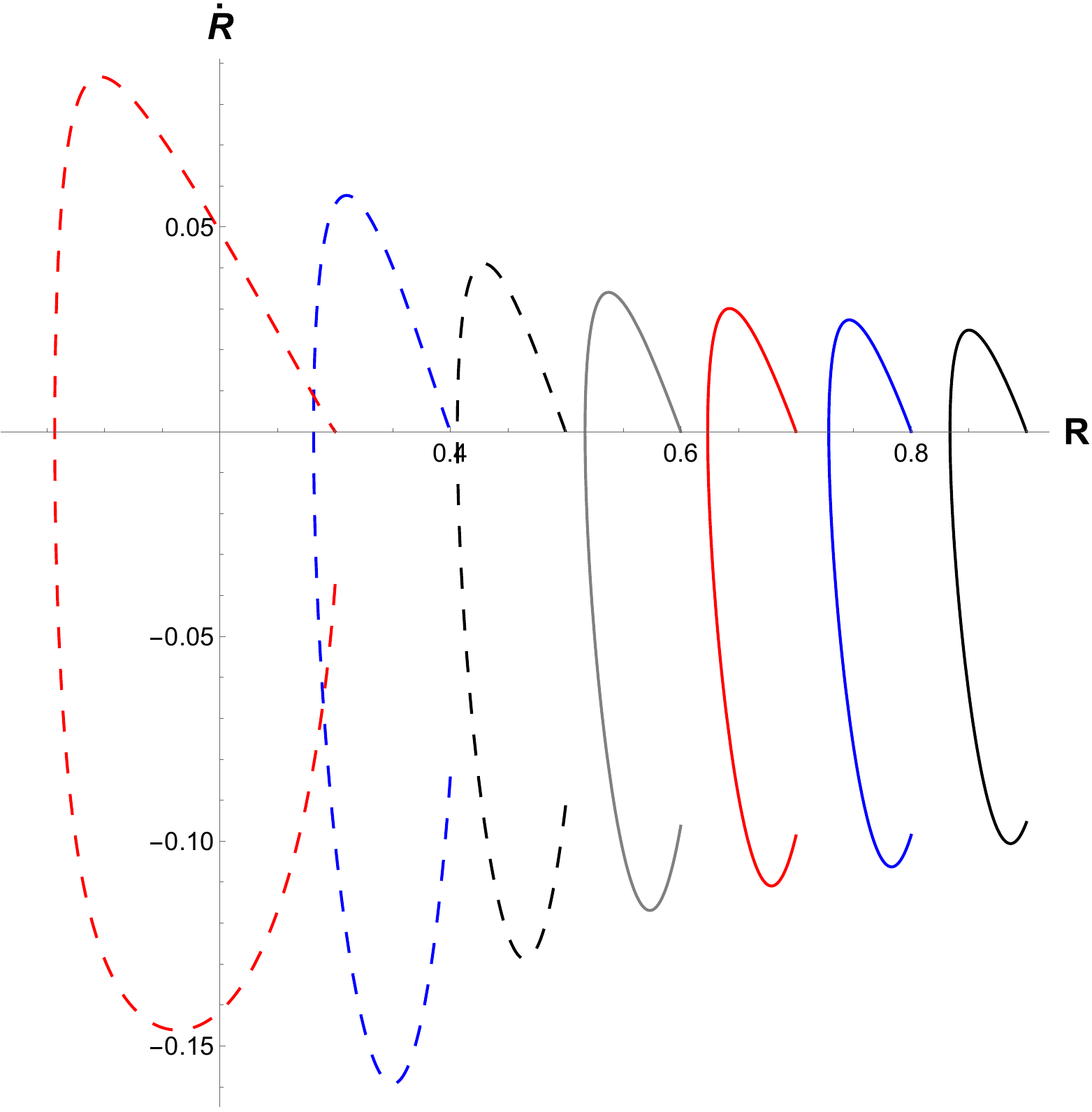}
		\caption{Behavior of the ratio of mass function over area radius (left panel) and $(\dot{R},R)$ curve for the same values of model parameters as of Fig.~(\ref{fig1}).}\label{fig3}
	\end{center}
\end{figure}
\begin{figure}
	\begin{center}
		\includegraphics[width=7.1cm]{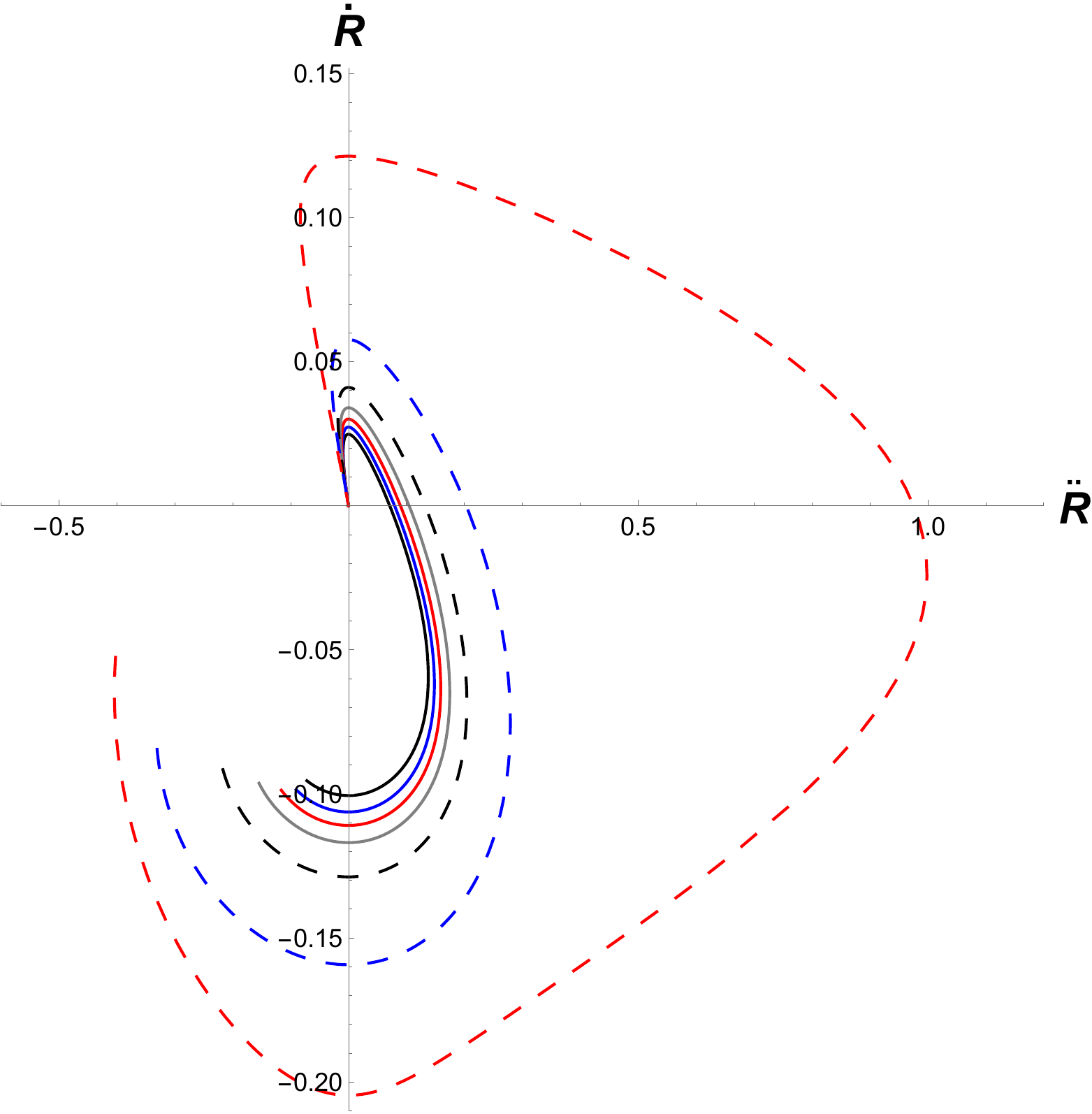}
		\includegraphics[width=7.7cm]{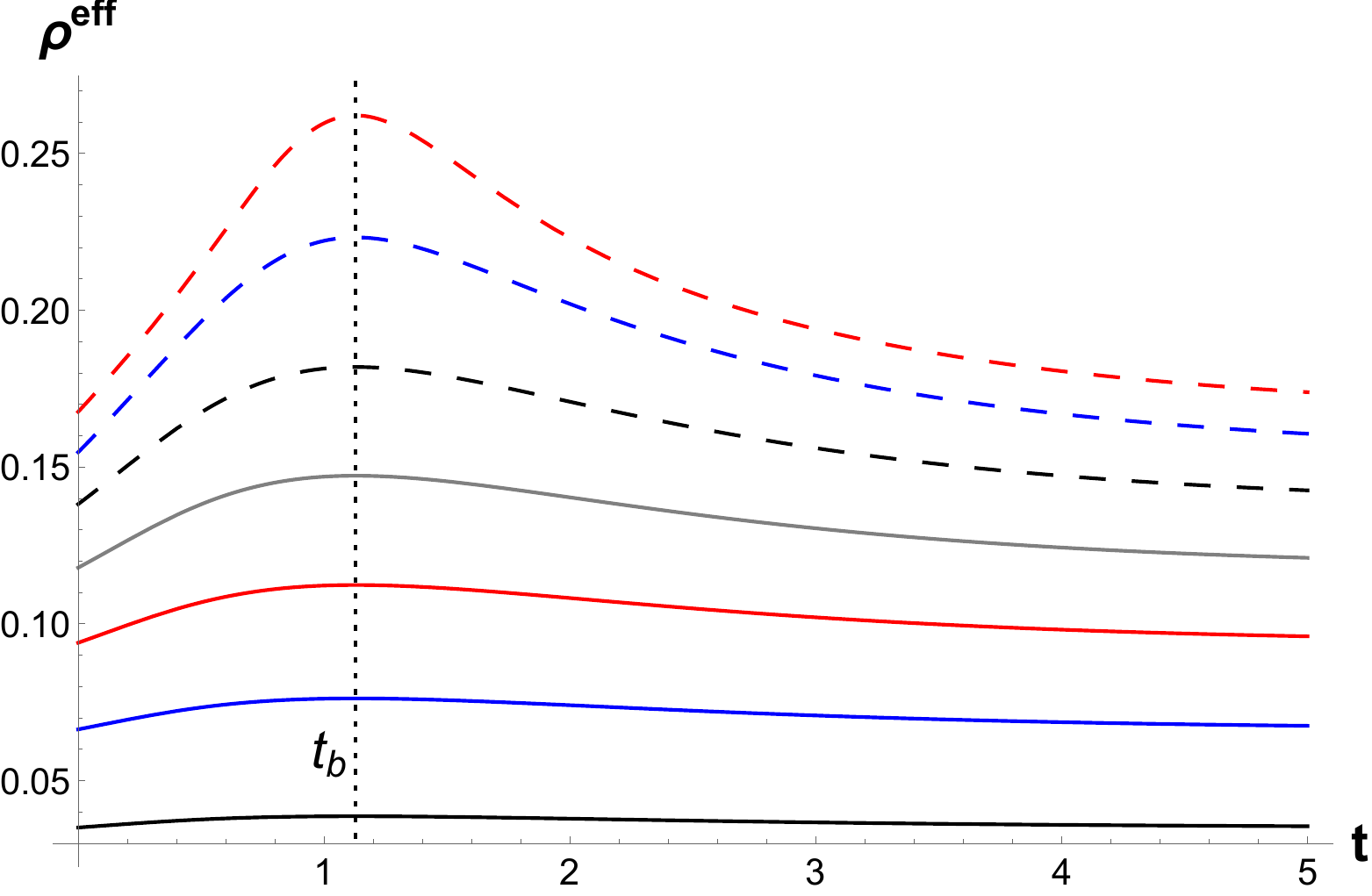}
		\caption{Behavior of the $(\dot{R},\ddot{R})$ curve (left panel) and the effective energy density (right panel) for the same values of model parameters as of Fig.~(\ref{fig1}).}\label{fig4}
	\end{center}
\end{figure}

An important issue that must be considered in every collapse scenario is to check whether the singular or nonsingular events that form as the collapse outcomes are necessarily trapped behind the horizon or visible to the external Universe. This can be pursued by studying the formation or otherwise of trapped surfaces during the dynamical evolution of the collapse process. Physically, the formation of trapped surfaces in a spherically symmetric gravitational collapse can be considered as, the amount of mass-energy accumulated within a given area radius of the collapsing cloud that the gravitational pull of which actually decides whether a two-sphere is trapped or not~\cite{MSEE}. Subsequently, during the dynamical process of the collapse, if it happens that $F(t,r)>R(t,r)$, then, trapped surface formation takes place and consequently the two-sphere is trapped, otherwise it is not~\cite{COUNTERCCC,Joshibook},\cite{trapped}. The left panel in Fig.~(\ref{fig3}) shows the behavior of the ratio $F/R$ during the collapse process where we observe that this ratio begins its evolution from $F/R<1$ at $t=0$ and remains less than unity at later times. This implies that no trapping of light occurs at the onset or during the dynamical evolution of the collapsing object, either at contracting or expanding phases. This result can be further verified considering the behavior of $\dot{R}$ against $R$, see the right panel of Fig.~(\ref{fig3}).
As the $(\dot{R},R)$ curve shows, the speed of collapse begins from negative values for each matter shell reaching a maximum value
in negative direction and vanishes at the bounce time. It then assumes a maximum value in the post-bounce regime and tends to zero at later times.
Hence, the speed of collapse is bounded during the dynamical evolution of the object in the collapsing phase and also in the expanding phase that follows after the bounce. Such a behavior is in contrast to the singular models where the speed of collapse diverges in the limit of approach to the shell-focusing singularity. As a simple example for this argument one may consider a marginally bound collapse of an inhomogeneous dust fluid for which the area radius is given as~\cite{joshsingh}
\be\label{arearad}
R(t,r)=\left(r^{\f{3}{2}}-\f{3}{2}\sqrt{F(r)}t\right)^{\f{2}{3}},~~~~~~~~\dot{R}(t,r)=-\sqrt{\f{F(r)}{R(t,r)}}.
\ee
The area radius for each shell vanishes at the singularity time which is given by $t_{\rm s}(r)=2r^{\f{3}{2}}/3\sqrt{F(r)}$.
It is therefore seen that $\dot{R}(t,r)\rightarrow-\infty$ as $t\rightarrow t_{\rm s}(r)$. Since $\dot{R}=0$ is not reached even asymptotically, the collapse then does not cease and all the matter shells are crushed into zero size at the final shell-focusing singularity. The resulting singularity may be either hidden by the event horizon of a black hole or be visible i.e., formation of a naked singularity~\cite{joshdwi}. The left panel in Fig.~(\ref{fig4}) shows the behavior of the curve $(\dot{R},\ddot{R})$ from which, as discussed earlier, we can firstly find the four phases of evolution of the collapsing body. We can further deduce that if the final outcome of the collapse process is to be a static, regular configuration then the collapse will eventually halt, reaching zero velocity and acceleration. As the figure shows, at later times in the post-bounce regime we have $\dot{R}=\ddot{R}=0$, which means that, a static configuration is achieved as the end-product of the collapse setting, see also~\cite{equijoshi} for further discussions.
\par 
Finally, in order to examine the fulfillment of the WEC we consider the first part of Eq.~(\ref{eqsfmass}) along with Eqs.~(\ref{EFFEMTw1}) and (\ref{Rasdust}) which provide us with the following expression for effective energy density at the beginning of the collapse
\be\label{rhoeffonset}
\rho^{\rm eff}(0,r)=\f{4}{3}(1+w_\theta)\rho(0,r)=\f{4\rho_0(1+w_\theta)(1-5w_\theta)^2}{243\kappa w_\theta^2}\left[1-\left(\f{r}{r_{\rm b}}\right)^n\right],~~~~~~~w_\theta\neq0.
\ee
From the above expression we find that the initial values of the effective energy density and the fluid energy density are positive provided that $\rho_0>0~\land w_\theta>-1$, as required by the regularity of initial conditions~\cite{COUNTERCCC}. We note that the lower bound on EoS parameter in tangential direction is given as $w_\theta>1/5$ which is demanded by positivity of the mass function. As the collapse proceeds, the energy density of the fluid and the effective one increase until reaching the bounce point where both densities assume maximum positive values, see the right panel of Fig.~(\ref{fig4}). This can also be seen through the first part of Eq.~(\ref{eqsfmass}) and the behavior of $R$ and $R^\prime$ for $t\leq t_{\rm b}$. In the post-bounce regime we observe that the effective energy density decreases to finite positive values, hence, it behaves regularly throughout the dynamical evolution of the collapse scenario. We note that the closer to the center of the cloud the matter shell the larger the associated energy density assumes the maximum values. From Eqs.~(\ref{WEC}) and (\ref{effemtwec}) we figure out that the same arguments hold for the radial and tangential components of WEC which are obtained as
\bea\label{radtanwec}
\rho^{\rm eff}\!\!+p^{\rm eff}_r=\rho+p_r=\f{4}{3}(1+w_\theta)\rho,~~~~~~~~~~\rho^{\rm eff}\!\!+p^{\rm eff}_\theta=\rho+p_\theta=(1+w_\theta)\rho.
\eea
\vspace{2cm}
\section{concluding remarks}\label{sec4}
In the present work, the collapse process of an inhomogeneous fluid was considered in Rastall gravity and it was shown that collapsing configuration evolves to a nonsingular bouncing scenario in which an expanding phase follows the contracting phase after the bounce event. This regular outcome is possible by setting the Rastall parameter in such a way that the effective pressure of the fluid in radial direction vanishes. Since this parameter can be regarded as a measure of the strength of non-minimal coupling between matter and geometry, one then may intuitively imagine that the mutual interaction between geometry and matter sources could avoid formation of shell-focusing singularities. We therefore observed that the collapsing body experiences four phases in the whole scenario without ever formation of shell-crossing singularities and achieves an equilibrium configuration where the speed as well as the acceleration of the matter shells tend to zero at the end of the expanding regime. Some points that beg more considerations are as follows: $i)$ while in nonsingular collapse scenarios the WEC is usually violated, especially in quantum corrected collapse models~\cite{quantuminhom1},\cite{quantumcorcol}, this condition holds for the present nonsingular solutions. $ii)$ The formation of spacetime singularities due to gravitational collapse in classical GR is widely regarded as inevitable under certain conditions, as established by the Hawking-Penrose singularity theorems~\cite{HAWPENST}. Though the observational and physical nature of a spacetime singularity remains an open question, a variety of singular models in GR have been reported that deal with the nature of the singularities. Some studies suggest that inhomogeneous or anisotropic collapse (e.g., in Lemaitre-Tolman-Bondi models) can lead to naked singularities (regions of extreme gravity that are visible to the observers in the Universe)~\cite{ColReview}. The formation of such objects contradicts the CCC, which states that singularities must always be hidden by a black hole event horizon therefore causally disconnected from the outside regions of the spacetime. Among the works on singular collapse scenarios we can quote: Tolman-Bondi dust collapse~\cite{JoshiDw1993}, inhomogeneous perfect fluid models~\cite{inpfcol,genersing0}, self-similar gravitational collapse models~\cite{selfsimcol}, role of initial data in final fate of dust collapse~\cite{singhjoshi1994},\cite{joshsingh}, singular collapse models in Vaidya~\cite{JoshiDw19930},\cite{nakedvaidya} and Szekeres~\cite{sczsing} spacetimes and other works~\cite{COUNTERCCC,Joshibook}, see also~\cite{shellcr21},\cite{genersing0},\cite{genersing} for genericity and stability of naked singularities and black holes that arise as the end-states of a complete gravitational collapse. Regarding these studies, one may argue that in the context of GR formation of spacetime singularities (either naked or dressed) as the collapse end products is generic, a scenario which is in contrast to the present nonsingular solutions in Rastall gravity. $iii)$ The equilibrium configuration that forms in the late stages of the post-bounce regime may be exposed to spherical or non-spherical\footnote{Non-spherical perturbations in stellar collapse to a black hole is one of the most promising sources for gravitational waves detectors~\cite{Harada2003}.} perturbations or even perturbations due to variations in the EoS parameters of the fluid. Therefore, the resulting equilibrium state can become unstable if these perturbations grow over time so that the whole matter distribution enters another phase of collapse or possibly undergoes oscillatory contracting and expanding phases~\cite{Harada2004}. We would like to conclude by saying that other types of solutions can be also achieved where the matter distribution starts to expand instead of undergoing a collapse process. These type of solutions may be useful for inhomogeneous cosmological setups~\cite{inhomcos} and possibly avoidance of the initial singularity of the Universe~\cite{nonsingcos}~\cite{bigb}. However, dealing with these solutions is beyond the scope of the present study. 
\par
\vspace{0.5cm}
{\bf Data Availability Statement}: The data that support the findings of this study are available from the corresponding author, [AJ], upon reasonable request.


\begin{thebibliography}{99}
\bibitem{HAWPENST} S. W. Hawking G. F. R and Ellis, {\it The Large Scale Structure of Space-Time,} Cambridge University Press (1973).
\bibitem{COUNTERCCC} P. S. Joshi, {\it Global Aspects in Gravitation and Cosmology,} Oxford University Press, Oxford (1993).
\bibitem{Joshibook} P. S. Joshi, {\it Gravitational Collapse and Space-Time Singularities,} Cambridge University Press, Cambridge, (2007).
\bibitem{CCCREF} R. Penrose, Nuovo Cimento {\bf 1} 252 (1969); Gen. Rel. Grav. {\bf 34} 1141 (2002);\\ Y. C. Ong, Int. J. Mod. Phys. A {\bf 35}, 2030007 (2020).
\bibitem{ColReview} P. S. Joshi and D. Malafarina, Int. J. Mod. Phys. D {\bf 20} 2641 (2011).
\bibitem{problemssin} R. Geroch, Annals of Phys. {\bf 48}, 526 (1968);\\ S. Hawking, Phys. Rev. D {\bf 14}, 2460 (1976);\\ G. F. R. Ellis and B. G. Schmidt, Gen. Relativ. Gravit. {\bf 8} 915 (1977);\\J. M. M. Senovilla and D. Garfinkle Class. Quantum Grav. {\bf 32} 124008 (2015).
\bibitem{problemssin1} C. J. S. Clarke, {\it The Analysis of Space-Time Singularities,} United Kingdom: Cambridge University Press (1993).
\bibitem{problemssin2} J. Earman, {\it Bangs, Crunches, Whimpers, and Shrieks: Singularities and Acausalities in Relativistic Spacetimes,} United Kingdom: Oxford University Press (1995).
\bibitem{modfssing} S. Nojiri and S. D. Odintsov, Phys. Rep. {\bf 505} 59 (2011);\\ T. Clifton, P. G. Ferreira, A. Padilla, C. Skordis, Phys. Rep., {\bf 513} 1 (2012);\\ K. Bamba, S. Nojiri, S. D. Odintsov, Phys. Lett. B {\bf 698}, 451 (2011);\\ G. Brando, F. T. Falciano, L. F. Guimaraes, Phys. Rev.
D {\bf 98}, 044027 (2018);\\ A. H. Ziaie, P. V. Moniz, A. Ranjbar, H. R. Sepangi, Eur. Phys. J. C {\bf 74}, 3154 (2014);\\ M. Hashemi, S. Jalalzadeh, A. H. Ziaie, Eur. Phys. J. C {\bf 75}, 53 (2015);\\ C. Bambi, D. Malafarina, A. Marciano, L. Modesto, Phys. Lett. B {\bf 734}, 27 (2014).
\bibitem{modfssingcomp} S. Capozziello and M. De Laurentis, Phys. Rep. {\bf 509}, 167 (2011);\\S. Nojiri and S. D. Odintsov, Phys. Rep., {\bf 505}, 59 (2011);\\ D. Malafarina and P. S. Joshi, Eur. Phys. J. C {\bf 75}, 596 (2015);\\ H. O. Silva, A. Maselli, M. Minamitsuji, E. Berti, Int. J Mod. Phys. D {\bf 25}, 1641006 (2016);\\A. C.-Turrion, A. de la C.-Dombriz, A. Dobado, Phys. Rev. D {\bf 105}, 084060 (2022);\\A. Zahra, S. A., Mardan, M. B. Riaz, R. Manzoor, I. Noureen, Eur. Phys. J. Plus {\bf 139}, 676 (2024).
\bibitem{stfarsin} V. Faraoni, {\it Cosmology in Scalar-Tensor Gravity,} Springer (2004).
\bibitem{modfssing1} A. H. Ziaie, H. Shabani, H. Moradpour, Eur. Phys. J. Plus {\bf 139}, 148 (2024).
\bibitem{branesing} R. Maartens and K. Koyama, Liv. Rev. Relativ. {\bf 13}, 5 (2010).
\bibitem{qsingavoid} M. Bojowald, Phys. Rev. Lett. {\bf 86}, 5227 (2001);\\ A. Ashtekar and P. Singh, Class. Quantum Grav. {\bf 28} 213001 (2011).
\bibitem{qsingavoid1} M. Bojowald, R. Goswami, R. Maartens, P. Singh, Phys. Rev. Lett. {\bf 95}, 091302 (2005);\\R. Goswami, P. S. Joshi, P. Singh, Phys. Rev. Lett. {\bf 96}, 031302 (2006); J. Ziprick and G. Kunstatter, Phys. Rev. D {\bf 82}, 044031 (2010);\\B.-F. Li and P. Singh, {\it Loop Quantum Cosmology: Physics of Singularity Resolution and Its Implications.} In: C. Bambi, L., Modesto, I., Shapiro, (eds) Handbook of Quantum Gravity, Springer, Singapore (2024);\\H. Singh, M. K. Nandy, Eur. Phys. J. C {\bf 84}, 700 (2024);\\M. Bojowald, G. M. Paily, Class. Quantum Grav. {\bf 29}, 242002 (2012);\\C. Kiefer, T. Schmitz, Phys. Rev. D {\bf 99}, 126010 (2019).
\bibitem{exocol} C. Rovelli and F. Vidotto, Int. J Mod. Phys. D {\bf 23}, 1442026 (2014); C. Rovelli and F. Vidotto, arXiv:2407.09584 [gr-qc];\\ K. Jusufi, Universe, {\bf 9}, 41 (2023).
\bibitem{quantuminhom} Y. Liu, D. Malafarina, L. Modesto, C. Bambi, Phys. Rev. D {\bf 90}, 044040 (2014).\\
A. Parvizi, T. Pawlowski, Y. Tavakoli, J. Lewandowski, Phys. Rev. D {\bf 105}, 086002 (2022);\\
J. Marto, Y. Tavakoli, P. V. Moniz, Int. J. Mod. Phys. D {\bf 24}, 1550025 (2015);\\
Y. Tavakoli, J. Marto, A. Dapor, Int. J. Mod. Phys. D {\bf 23} 1450061 (2014).
\bibitem{qcolbounce} C. Bambi, D. Malafarina and L. Modesto, Phys. Rev. D {\bf 88}, 044009 (2013);\\C. Bambi, D. Malafarina, L. Modesto, Eur. Phys. J. C {\bf 74}, 2767 (2014);\\P. S. Joshi, D. Malafarina, (eds.) {\it New Frontiers in Gravitational Collapse and Spacetime Singularities,} Germany:
Springer Nature Singapore, Imprint: Springer (2024).
\bibitem{nonsingcos} H. Shabani and A. H. Ziaie, Eur. Phys. J. C {\bf 78}, 397 (2018);\\J. K. Singh, H. Balhara, K., Bamba, J. Jena J. High Energ. Phys. 2023, 191 (2023);\\O. Trivedi, Symmetry {\bf 16}, 298 (2024);\\ M. Zubair, M. Farooq, E. Gudekli, Int. J. Geom. Meth. Mod. Phys. {\bf 19}, 2250135 (2022);\\M. Caruana, G., Farrugia, J. Levi Said, Eur. Phys. J. C {\bf 80}, 640 (2020);\\ M. Sharif, M. Z. Gul, N. Fatima, New Astronomy {\bf 109}, 102211 (2024);\\M. Z. Gul, M. Sharif, S. Shabbir, Eur. Phys. J. C {\bf 84}, 802 (2024);\\ M. Sharif, M. Z. Gul, I. Hashim, Eur. Phys. J. C {\bf 84}, 1094 (2024);\\M. Sharif, M. Z. Gul, N. Fatima, Phys. Dark Univ. {\bf 48}, 101839 (2025).
\bibitem{od1} S. Nojiri, S. D. Odintsov, Phys. Lett. B {\bf599}, 137 (2004).
\bibitem{od2} G. Allemandi, A. Borowiec, M. Francaviglia, S. D. Odintsov, Phys. Rev. D {\bf72}, 063505 (2005).
\bibitem{cmc} T. Koivisto, Class. Quant. Grav. {\bf23}, 4289 (2006).
\bibitem{cmc1} O. Bertolami, C. G. Boehmer, T. Harko, F. S. N. Lobo, Phys. Rev. D {\bf75}, 104016 (2007).
\bibitem{cmc2} T. Harko, F. S. N. Lobo, Galaxies, {\bf2}, 410 (2014).
\bibitem{rastall} P. Rastall, Phys. Rev. D {\bf 6}, 3357 (1972);\\ Can. J. Phys. {\bf 54}, 66 (1976).
\bibitem{pppcos} L. Parker, Phys. Rev. D {\bf 3}, 346 (1971); {\bf 3}, 2546 (1971);\\ L. H. Ford, Phys. Rev. D {\bf 35}, 2955 (1987);\\
G. W. Gibbons and S. W. Hawking, Phys. Rev. D {\bf 15}, 2738 (1977);\\ N. D. Birrell and P. C. W. Davies, {\it Quantum Fields in Curved Space,} Cambridge University Press, Cambridge (1982).
\bibitem{qusgrf} S. Nojiri, S. D. Odintsov, Phys. Lett. B {\bf 599}, 137 (2004);\\G. Allemandi, A. Borowiec, M. Francaviglia, S. D. Odintsov, Phys. Rev. D {\bf 72}, 063505 (2005); \\T. Koivisto, Class. Quant. Grav. {\bf 23}, 4289 (2006); O. Bertolami, C. G. Boehmer, T. Harko, F. S. N. Lobo, Phys. Rev. D {\bf 75}, 104016 (2007);\\ T. Harko, F. S. N. Lobo, Galaxies, {\bf 2}, 410 (2014);\\ M. F. A. R. Sakti, A. Suroso, A. Sulaksono, F. P. Zen, Phys. Dark Univ. {\bf 35}, 100974 (2022).
\bibitem{genras} H. Moradpour, Y. Heydarzade, F. Darabi, I. G. Salako, Eur. Phys. J. C {\bf 77}, 259 (2017).
\bibitem{dmy} F. Darabi, H. Moradpour, I. Licata, Y. Heydarzade, C. Corda, Eur. Phys. J. C {\bf 78}, 25 (2018).
\bibitem{FESRASPLB} A. S. Al-Rawaf and M. O. Taha, Phys. Lett. B {\bf 366} 69 (1996).
\bibitem{Effemtras} A. S. AI-Rawaf and M. O. Taha, Gen. Relative. Gravit., {\bf 28} 935 (1996);\\ H. Moradpour, N. Sadeghnezhad, S. H. Hendi, Can. J. Phys, {\bf 95}, 1257 (2017).
\bibitem{validity1} A. S. Al-Rawaf and M. O. Taha, Phys. Lett. B {\bf 366} 69 (1996).
\bibitem{validity2} K. A. Bronnikov, J. C. Fabris, O. F. Piattella and E. C. Santos, Gen. Rel. Grav. {\bf 48}, 162 (2016).
\bibitem{colrast} A. H. Ziaie, H. Moradpour, S. Ghaffari, Phys. Lett. B {\bf 793}, 276 (2019).
\bibitem{ziataval} A. H. Ziaie and Y. Tavakoli, {\bf 532}, 2000064 (2020).
\bibitem{OSDM} S. Datt, Zs. f. Phys. {\bf 108}, 314 (1938);\\J. R. Oppenheimer and H. Snyder, Phys. Rev. {\bf 56}, 455 (1939).
\bibitem{colrast1} A. H. Ziaie, H. Moradpour, M. Mohammadi Sabet, Eur. Phys. J. Plus, {\bf 136}, 1085 (2021).
\bibitem{eqrasgr} M. Visser, Phys. Lett. B {\bf 782}, 83 (2018);\\A. Golovnev, Ann. Phys. {\bf 461}, 169580 (2024).
\bibitem{validras} R. Li , J. Wang , Z. Xu , X. Guo, MNRAS, {\bf 486}, 2407 (2019);\\O. Akarsu, N. Katirci, S. Kumar, R. C. Nunes, B. Ozturk, S. Sharma, Eur. Phys. J. C {\bf 80}, 1050 (2020);\\ J. M. Z. Pretel and C. E. Mota, Gen. Relativ. Gravit. {\bf 56}, 43 (2024);\\ Q. Sun, Y. Zhang, C.-H. Xie, Q.-Q. Li, Phys. Dark Univ. {\bf 46}, 101599 (2024);\\ J. A.-Moreno, K. Jacobo, S. Arteaga, M. A. G.-Aspeitia, A. H.-Almada, Class. Quantum Grav. {\bf 41}, 065003 (2024).
\bibitem{glensras} A. M. M. Abdel-Rahman, Astrophys. Space Sci. {\bf 278}, 385 (2001);\\A. M. M. Abdel-Rahman and M. H. A. Hashim, Astrophys. Space Sci. {\bf 298}, 519 (2005).
\bibitem{fluidras} J. Chagoya, J. C. L.-Dominguez, C. Ortiz, Class. Quantum Grav. {\bf 40}, 075005 (2023).
\bibitem{olive2015} A. M. Oliveira, H. E. S. Velten, J. C. Fabris,  L. Casarini,Phys. Rev. D {\bf92}, 044020 (2015);\\ W. E. Hanafy, ApJ {\bf940},51 (2022);\\G. G. L. Nashed, W. E. Hanafy, Eur. Phys. J. C{\bf 82},679 (2022);\\J. M. Z. Pretel, C. E Mota, Gen. Relativ. Gravit. {\bf 56}, 43 (2024).
\bibitem{singhjoshi1994} T. P. Singh, P. S. Joshi, Class. Quant. Grav. {\bf 13}, 559 (1996);\\ S. Jhingan, P. S. Joshi, T. P. Singh, Class. Quant. Grav. {\bf 13}, 3057 (1996);\\ I. H. Dwivedi and P. S. Joshi, Class. Quant. Grav. {\bf 14}, 1223 (1997).
\bibitem{JoshiDw1993} P. S. Joshi, I. H. Dwivedi, Phys. Rev. D {\bf 47}, 5357 (1993).
\bibitem{JoshiDw19930} I. H. Dwivedi and P. S. Joshi, Class. Quantum Grav. {\bf 6}, 1599 (1989).
\bibitem{JoshiDw19931} P. S. Joshi, N. Dadhich, R. Maartens, Phys. Rev. D {\bf 65}, 101501 (2002);\\R. V. Saraykar and S. H. Ghate, Class. Quantum Grav. {\bf 16}, 281 (1999);\\T. Harada, H. Iguchi, K.-ichi Nakao, Prog. Theor. Phys. {\bf 107}, 449 (2002).
\bibitem{sczsing} P. S. Joshi, A. Krolak, Class. Quantum Grav. {\bf 13}, 3069 (1996).
\bibitem{quantuminhom1} V. Husain, Adv. Sci. Lett. {\bf 2}, 214 (2009);\\ D. Malafarina, Universe {\bf 3}, 48 (2017);\\S. G. Choudhury and S. Chakrabarti, JCAP 01 (2024) 007.
\bibitem{luzmenazia} P. Luz, F. C. Mena, A. H. Ziaie, Class. Quant. Grav. {\bf 36}, 015003 (2019).
\bibitem{weakfield} H. Moradpour, I. G. Salako, Adv. High Energy Phys. {\bf 2016}, 2016, 3492796;\\ M. Capone, V. F. Cardone, M. L. Ruggiero, J. Phys.: Conf. Ser., {\bf 222}, 012012 (2010). 
\bibitem{Stef2003} H. Stephani, D. Kramer, M. MacCallum, C. Hoenselaers, E. Herlt, {\it Exact Solutions of Einstein's Field Equations,} Cambridge: Cambridge University Press (2003).
\bibitem{Wang1999} A. Wang and Y. Wu, Gen. Relativ. Gravit. {\bf 31} 107 (1999);\\ S. D. Maharaj, G. Govender and M. Govender, Gen. Relativ. Gravit. {\bf 44} 1089 (2012).
\bibitem{nakedvaidya} K. Lake, Phys. Rev. D {\bf 43}, 1416 (1991);\\ I. H. Dwivedi and P. S. Joshi, Class. Quantum Grav. {\bf 8}, 1339 (1991);\\S. M. Wagh, S. D. Maharaj, Gen. Rel. Grav. {\bf 31}, 975 (1999);\\ S. G. Ghosh, Phys. Rev. D {\bf 62}, 127505 (2000);\\S. G. Ghosh, N. Dadhich, Phys. Rev. D {\bf 64}, 047501 (2001);\\S. Jhingan, N. Dadhich, P. S. Joshi, Phys. Rev. D {\bf 63}, 044010 (2001);\\T. Harko, Phys. Rev. D {\bf 68}, 064005 (2003);\\J. Wheeler, Class. Quantum Grav. {\bf 39}, 197001 (2022).
\bibitem{MSE} C. W. Misner and D. H. Sharp, Phys. Rev. {\bf 136} B571 (1964);\\D. Bak and S. J. Rey, Class. Quantum Grav. {\bf 17} L83 (2000);\\
T. Koike, H. Onozawa and M. Siino, arXiv:gr-qc/9312012.
\bibitem{polyeos} U. S. Nilsson and C. Uggla, Ann. Phys. {\bf 286}, 292 (2001).
\bibitem{polyeos1} L. Rezzolla and O. Zanotti, {\it Relativistic Hydrodynamics,} United Kingdom, Oxford (2013).
\bibitem{quadeos} T. Feroze, A. A. Siddiqui, Gen. Relativ. Gravit. {\bf 43}, 1025 (2011);\\S. D. Maharaj and P. M. Takisa, Gen. Relativ. Gravit. {\bf 44}, 1419 (2012);\\J. M. Sunzu and A. V. Mathias, Indian J. Phys. {\bf 96}, 4059 (2022).
\bibitem{polchapeos} A. K. Prasad, J. Kumar, A. Sarkar, Gen. Relativ. Gravit. {\bf 53}, 108 (2021);\\J. M. Sunzu, A. V. Mathias, Indian J. Phys. {\bf 97}, 687 (2023);\\ D. Bhattacharjee, P. K. Chattopadhyay, Eur. Phys. J. C {\bf 84}, 77 (2024).
\bibitem{vandereos} S. Capozziello, S. De Martino, M. Falanga, Phys. Lett. A {\bf 299}, 494 (2002);\\F. S. N. Lobo, Phys. Rev. D {\bf 75}, 024023 (2007);\\R. C. S. Jantsch, M. H. B. Christmann, G. M. Kremer, Int. J. Mod. Phys. D {\bf 25}, 1650031 (2016).
\bibitem{logeos} P. -H. Chavanis, Eur. Phys. J. Plus {\bf 130}, 130 (2015);\\ H. Benaoum, P.-H. Chavanis, H. Quevedo, Universe {\bf 2022}, 8, 468.
\bibitem{lucadarkeos} L. Amendola and S. Tsujikawa, {\it Dark Energy: Theory and Observations} Cambridge University Press (2010).
\bibitem{vangoozi} L. A. Escamilla, W. Giare, E. D. Valentino, R. C. Nunes, S. Vagnozzi, JCAP 05 (2024) 091.
\bibitem{quinteeos} V. Sahni and A. Starobinsky, Int. J. Mod. Phys. D {\bf 09}, 373 (2000).
\bibitem{phantomeos} Y.-P. Teng, W. Lee, K.-W. Ng, Phys. Rev. D {\bf 104}, 083519 (2021).
\bibitem{thermoras} I. P. Lobo, H. Moradpour, J. P. Morais Graca, I. G. Salako, Int. J. Mod. Phys. D {\bf 27}, 1850069 (2018).
\bibitem{darkrastall} J. C. Fabris, O. F. Piattella, D. C. Rodrigues, C. E. M. Batista, M. H. Dauda, Int. J. M. Phys. Conf. Series, {\bf 18}, 67 (2012);\\ M. Capone, V. F. Cardone, M. L. Ruggiero, J. Phys.: Conf. Ser. {\bf 222}, 012012 (2010);\\ H. Shabani, H. Moradpour, A. H. Ziaie, Phys. Dark Univ. {\bf 36} 101047 (2022).
\bibitem{observgama} M. Tang , Z. Xu, J. Wang, Chinese Phys. C {\bf 44}, 085104 (2020);\\C. E. M. Batista, J. C. Fabris, O. F. Piattella, A. M. V.-Toribio, Eur. Phys. J. C {\bf 73}, 2425 (2013).
\bibitem{inhomcos} A. Krasinski, {\it Inhomogeneous Cosmological Models,} United Kingdom: Cambridge University Press (2006).
\bibitem{shellcr1} P. S. Joshi, R. V. Saraykar, Int. J. Mod. Phys. D {\bf 22}, 1350027 (2013).
\bibitem{shellcr} S. M. C. V. Goncalves, Phys. Rev. D {\bf 63}, 124017 (2001);\\ K. Bolejko and P. Lasky, Mon. Not. R. Astro. Soc.: Letters, {\bf 391}, L59 (2008).
\bibitem{ellissing} G. F. R. Ellis, B. G. Schmidt, Gen. Relat. Gravit. {\bf 8}, 915 (1977);\\ C. B. Collins and G. F. R. Ellis, Phys. Rep. {\bf 56}, 67 (1979).
\bibitem{shellcr2} B. C. Nolan, Class. Quantum Grav. {\bf20}, 575 (2003);\\K. N. Solanki, K. Mosani, O. Deshpande, P. S Joshi, Class. Quantum Grav. {\bf 41}, 165012 (2024).
\bibitem{shellcr21} P. S. Joshi, D. Malafarina, R. V. Saraykar, Int. J. Mod. Phys. D {\bf 21}, 1250066 (2012).
\bibitem{MSEE} S. A. Hayward, Phys. Rev. D {\bf 49} 831 (1994); Phys. Rev. D {\bf 49} 6467 (1994); Phys. Rev. D {\bf 53} 1938 (1994).
%
\bibitem{trapped} P. S. Joshi and R. Goswami, Class. Quantum Grav. {\bf 24}, 2917 (2007).
\bibitem{joshsingh} P. S. Joshi, T. P. Singh, Phys. Rev. D {\bf 51}, 6778 (1995).
\bibitem{joshdwi} P. S. Joshi, I. H. Dwivedi, Class. Quantum Gravity {\bf 16}, 41 (1999).
\bibitem{equijoshi} P. S. Joshi, D. Malafarina, R. Narayan, Class. Quantum Grav. {\bf 28} 235018 (2011);\\D. Malafarina, P. S. Joshi, Eur. Phys. J. C {\bf 75}, 596 (2015).
\bibitem{quantumcorcol} B. K. Tippett and V. Husain, Phys. Rev. D {\bf 84}, 104031 (2011);\\C. Barcelo, R. C.-Rubio, L. J. Garay, Universe {\bf 2016}, 2(2), 7;\\X. Calmet, R. Casadio, F. Kuipers, Phys. Lett. B
{\bf 807}, 135605 (2020);\\ E. Bittencourt, A. G. Cesar, J. P. Pereira, JCAP {\bf 12}, 037 (2023).
\bibitem{inpfcol} K. Mosani, D. Dey, P. S. Joshi, Phys. Rev. D {\bf 101}, 044052 (2020).
\bibitem{genersing0} S. Satin, D. Malafarina, P. S. Joshi,	Int. J. Mod. Phys. D {\bf 25}, 1650023 (2016).
\bibitem{selfsimcol} P. S. Joshi, I. H. Dwivedi, Lett. Math. Phys. {\bf 27}, 235 (1993).
\bibitem{genersing} P. S. Joshi, Pramana J. Phys. {\bf 69}, 119 (2007).
\bibitem{Harada2003} T. Harada, H. Iguchi, M. Shibata, Phys. Rev. D {\bf 68}, 024002 (2003);\\ C. L. Fryer, K. C. B. New, Living Rev. Relativ. {\bf 14}, 1 (2011).
\bibitem{Harada2004} T. Harada, Pramana J. Phys. {\bf 63}, 741 (2004);\\
J. M. Z. Pretel, Eur. Phys. J. C {\bf 80}, 726 (2020).
\bibitem{bigb} A. Ijjas and P. J. Steinhardt, Class. Quantum Grav. {\bf 35}, 135004 (2018);\\I. Banerjee, T., Paul, S. SenGupta, Gen. Relativ. Gravit. {\bf 54}, 119 (2022);\\L. Fabbri, Universe {\bf 8}, 51 (2022);\\ M. Z. Gul, M. Sharif, S. Shabbir, Eur. Phys. J. C {\bf 84}, 802 (2024).

\end{thebibliography}
\end{document}